\documentclass[10pt,letterpaper]{article}
\usepackage[top=1in,left=1.25in,footskip=0.75in]{geometry}

\usepackage{changepage}

\usepackage[utf8x]{inputenc}

\usepackage{textcomp,marvosym}

\usepackage{fixltx2e}

\usepackage{amsmath,amssymb}

\usepackage{cite}

\usepackage{nameref,hyperref}

\usepackage[right]{lineno}

\usepackage[pdftex]{graphicx}

\usepackage{microtype}
\DisableLigatures[f]{encoding = *, family = * }

\raggedright
\usepackage[aboveskip=1pt,labelfont=bf,labelsep=period,justification=raggedright,singlelinecheck=off]{caption}

\makeatletter
\renewcommand{\@biblabel}[1]{\quad#1.}
\makeatother

\date{}

\newcommand{\beginsupplement}{%
        \setcounter{table}{0}
        \renewcommand{\thetable}{S\arabic{table}}%
        \setcounter{figure}{0}
        \renewcommand{\thefigure}{S\arabic{figure}}%
     }

\begin{document}
\vspace*{0.2in}

\begin{flushleft}
{\Large
\textbf{Individual Differences in Dynamic Functional Brain Connectivity Across the Human Lifespan}
}
\newline
\\
Elizabeth N. Davison \textsuperscript{1,*},
Benjamin O. Turner\textsuperscript{2},
Kimberly J. Schlesinger \textsuperscript{3},
Michael B. Miller\textsuperscript{2},
 Scott T. Grafton\textsuperscript{2},
Danielle S. Bassett\textsuperscript{4,5},
Jean M. Carlson\textsuperscript{3},
\\
\bigskip
\textbf{1}  Department of Mechanical and Aerospace Engineering, Princeton University, Princeton, New Jersey, United States of America 
\\
\textbf{2}  Department of Psychological and Brain Sciences, University of California, Santa Barbara, Santa Barbara, California, United States of America 
\\
\textbf{3}  Department of Physics, University of California, Santa Barbara, Santa Barbara, California, United States of America 
\\
\textbf{4}   Department of Bioengineering, University of Pennsylvania, Philadelphia, Pennsylvania, United States of America 
\\
\textbf{5}   Department of Electrical and Systems Engineering, University of Pennsylvania, Philadelphia, Pennsylvania, United States of America 
\\
\bigskip

* end@princeton.edu

\end{flushleft}
\section*{Abstract}
	Individual differences in brain functional networks may be related to complex personal identifiers, including health, age, and ability. Understanding and quantifying these differences is a necessary first step towards developing predictive methods derived from network topology. Here, we present a method to quantify individual differences in brain functional dynamics by applying hypergraph analysis, a method from dynamic network theory.  Using a summary metric derived from the hypergraph formalism---hypergraph cardinality---we investigate individual variations in two separate and complementary data sets. The first data set (``multi-task'') consists of 77 individuals engaging in four consecutive cognitive tasks. We observed that hypergraph cardinality exhibits variation across individuals while remaining consistent within individuals between tasks; moreover, one of the memory tasks evinced a marginally significant correspondence between hypergraph cardinality and age. This finding motivated a similar analysis of the second data set (``age-memory''), in which 95 individuals of varying ages performed a memory task with a similar structure to the multi-task memory task. With the increased age range in the age-memory data set, the correlation between hypergraph cardinality and age correspondence becomes significant. We discuss these results in the context of the well-known finding linking age with network structure, and suggest that age-related changes in brain function can be better understood by taking an integrative approach that incorporates information about the dynamics of functional interactions.

\section*{Author Summary}

Complex patterns of activity in each individual human brain generates the
unique range of thoughts and behaviors that person experiences. Individual
differences in ability, age, state of mind, and other characteristics are
tied to differences in brain activity, but determination of the exact
nature of these relationships has been limited by the intrinsic complexity
of the brain. Here, we apply dynamic network theory to quantify
fundamental features of individual neural activity.  We represent
functional connections between brain regions as a time varying network,
and then identify groups of these interactions that exhibit similar
behavior over time. The result of this construction is referred to as a
hypergraph, and each grouping within the hypergraph is called a
hyperedge.  We find that the number of these hyperedges in an individual's
hypergraph is a trait-like metric, with considerable variation across the
population of subjects, but remarkable consistency within each subject as
they perform different tasks. We find a significant correspondence between
this metric and the subject's age, indicating that the dynamics of
functional brain activity in older individuals tends to be more
dynamically segregated. This new insight into age-related changes in the
dynamics of cognitive processing expands our knowledge of the effects of
age on brain function and confirms our methods as promising
for quantifying and examining individual differences.


\section*{Introduction}
Functional connectivity (FC) analyses based on fMRI data are effective tools for quantifying and characterizing interactions between brain regions. Many approaches borrow methods from the field of graph theory, in which FC is used to build graphs that model the brain as a complex network, treating brain regions as nodes and using functional connections (pairs of nodes with significantly related BOLD signal dynamics) to determine the edge structure of the network \cite{bullmore2009complex, friston2011functional}. Individual differences in both underlying FC and the complex network structure resulting from graph theory approaches have been investigated for a variety of task states, developmental stages, and clinical diagnoses \cite{tavor2016task,zhang2010disease, greicius2008resting}.

Certain characteristics of FC have been found to vary consistently over the course of normal human aging. The loss of clear segmentation between neural systems is widely reported: many intrinsic functional connectivity networks in the brain tend to become less internally coherent with age, and the functional differences between these intrinsic networks generally become less pronounced \cite{dennis2014functional, contreras2015structural, sala2015reorganization}. These changes are most commonly reported in the default mode network (DMN) \cite{onoda2012decreased, tomasi2012aging, wang2012decoding, song2014age, ferreira2015aging, geerligs2015brain, ng_reduced_2016}, although they have also been observed in other networks, including those associated with higher cognitive functions \cite{onoda2012decreased, wang2012decoding, chan2014decreased, geerligs2015brain, ng_reduced_2016}. In addition, inter-network connectivity between the DMN and other regions of the brain has been found to increase, diminishing the ability to discriminate between networks based on FC \cite{ferreira2015aging,ng_reduced_2016}. There are some intrinsic functional networks, however, that show no changes or even increased intra-network conectivity with age, such as sensory networks\cite{tomasi2012aging, song2014age, geerligs2015brain}.


The bulk of studies on age-related changes and other individual differences in FC, including those that use methods from complex networks and graph theory to represent FC patterns, are performed using static FC analysis, which represents the similarities of brain region activity (or some other measure of concordance) aggregated across an entire data set. Here, we build upon recent advances in network science to study individual differences in human brain activity and behavior from a dynamic network science perspective \cite{holme2012temporal}. Dynamic functional connectivity (DFC) extends FC to examine how functional organization evolves over time \cite{hutchison2013dynamic, calhoun2014chronnectome}, allowing investigation of the changes in FC during the course of a cognitive task or scanning session. Efforts to probe the dynamics of functional brain networks have revealed that functional structure reconfigures over time in response to task demands \cite{doron2012dynamic, cohen2014quantifying, monti2014estimating, gonzalez-castillo_tracking_2015, davison_brain_2015} and spontaneously at rest \cite{hutchison2013dynamic, zalesky_time-resolved_2014}. DFC methods have also been used to inform understanding of individual differences related to aging. In particular, dynamic community structure was found to vary significantly with age \cite{Schlesinger2016age} and amplitude of low-frequency fluctuations of FC (ALFF-FC) was used to show age-dependent changes in the dynamics of interactions between networks \cite{qin2015predicting}. Both studies imply that functional dynamics should be considered when investigating how aging affects brain network organization.

Here, we use hypergraph analysis to examine individual differences in DFC network structure in fMRI data acquired as subjects perform cognitively demanding tasks. Compared to traditional graph theoretic techniques, hypergraphs address an existing methodological gap by characterizing not only activity, but also the co-evolution of activity over time. Hyperedges group connections that co-vary over both strong and weak interactions, thus enabling a more complete description of activity during both rest and cognitive tasks. Hypergraph methods extend standard graph methods to incorporate information about co-evolution of activity; whereas standard methods operate on the node-node connectivity matrix, hypergraph methods operate on the edge-edge connectivity matrix. The particular method we utilize here identifies groups of FC connections with statistically similar temporal profiles and links them into groups called hyperedges \cite{bassett_cross-linked_2014}. Standard FC characterizes interactions between pairs of brain regions and can be extended through DFC methods to capture the dynamics of those interactions. The groups of brain regions that comprise hyperedges are not necessarily strongly active or strongly interconnected brain regions. Rather, correlations in the dynamic connectivity of these regions are the defining characteristics that determine hyperedge structure. As a result, hypergraph analysis is able to identify groups of dynamic connections that change from strong to weak (or {\it vice versa}) cohesively together over time, providing complementary information to other DFC methods that focus on only the strongest node-node correlations, such as dynamic community detection \cite{bassett_robust_2013, bassett2013task, Schlesinger2016age}.

In previous work, we demonstrated that hyperedges discriminate between diverse task states in a group-level analysis of an fMRI data set spanning four tasks, which we refer to as the ``multi-task" data set \cite{davison_brain_2015}. We also observed notable variation in descriptive hypergraph measures across individuals. In this paper, we extend these results by developing and employing hypergraph measures that capture individual differences in functional brain dynamics to determine correspondences between dynamics and specific demographic and behavioral measures. In the multi-task data set, we find that hypergraph cardinality---the number of distinct hyperedges within a subject's hypergraph---exhibits marked variation across individuals. At the same time, we find this measure is consistent within individuals, across overall hypergraphs and those associated with specific tasks.

To elucidate the drivers of this striking variation in hypergraph metrics observed across subjects, we explore systematic relationships between hypergraph cardinality and individual difference measures spanning distinct domains such as demographics, cognitive strategy, and personality. In the multi-task data set, we find a suggestive relationship between hypergraph cardinality and participant age. This relationship is confirmed with an independent analysis of a data set with participants who range in age from 18 to 75, which we refer to as the ``age-memory'' data set. We report a strong positive relationship between age and hypergraph cardinality: older participants are significantly more likely to have a larger number of distinct hyperedges in their hypergraph. This agrees with the widely reported phenomenon of the loss of cohesion within intrinsic functional brain systems, because an increase in the number of distinct hyperedges linking various brain regions points to interconnections between functional groups evolving in time \cite{ferreira2015aging, ng_reduced_2016}. Thus, the hypergraph method agrees with previous descriptions of age-related brain changes, while capturing information about dynamics that adds a novel dimension to previous studies. This work further recommends the hypergraph as a useful tool in studying structure in dynamic functional connectivity.

\section*{Methods}
\subsection*{Ethics Statement}

Informed written consent was obtained from each participant prior to experimental sessions for the multi-task and age-memory experiments. All procedures were approved by the University of California, Santa Barbara Human Participants Committee.

\subsection*{Background and Multi-Task Methods}

\subsubsection*{Multi-Task Experimental Design}

Participants were scanned at rest (task-free) and while engaging in three distinct tasks designed to elicit distinct cognitive functions: an attention-demanding task, a memory task with lexical stimuli, and a memory task with face stimuli. Participants were instructed to lie still and look at a blank screen for the duration of the rest period. During the attention task, participants were instructed to attend to sequences of images on a screen and detect the presence or absence of a target stimulus in designated test displays. Prior to the test display, a cue arrow provided probabilistic information on whether and where the target stimulus might appear. The test display was flashed for approximately 50 ms, after which participants chose whether or not the target stimulus had been present. In both memory tasks, participants were presented with 180 previously examined stimuli and 180 novel stimuli and were asked to discriminate between the two. The memory tasks also included probabilistic cues indicating the probability that the stimulus was novel. For additional experimental details, see \cite{hermundstad_structural_2013} and \cite{aminoff_individual_2012}.

After completing the scans described above, the following individual difference measures were obtained for study participants: self-reported demographic information, self-reported state of mind (including physical and mental comfort) information, results from the Beck Depression Inventory II \cite{beck1988inventory}, tests for cognitive style (Santa Barbara Learning Style Questionnaire \cite{massa2006testing}, Object Spatial Imagery Questionnaire\cite{blajenkova2006object},  The Need for
Cognition Questionnaire \cite{cacioppo1982need}, Verbalizer-Visualizer Questionnaire \cite{richardson1977VVQ}, Card Rotation and Paper Folding Tests \cite{ekstrom1976manual}), personality tests (Big Five Inventory \cite{john1999big}  BIS/BAS scales\cite{carver1994behavioral}, and PANAS mood assessment \cite{watson1988development}). More individual difference measures were also collected, but do not match the individual difference measures collected from subjects in the age-memory study.

\subsubsection*{Image Acquisition and Processing}

The MRI data were acquired from 116 participants at the UCSB Brain Imaging Center using a phased array 3T Siemens TIM Trio with a 12 channel head coil. In addition to functional data, a three dimensional high-resolution T1-weighted structural image of the whole brain was obtained for each participant. Functional MRI data were collected from 116 healthy adult participants over the four states described above. Due to various sources of attrition, only 77 participants completed the functional scan and accompanying survey of detailed in \cite{aminoff_individual_2012}. The sampling period (TR) was 2 s for the rest and attention tasks and 2.5 s for both memory tasks  (TE = 30ms, FA = 90).

The functional data is parcellated into regions using a ``hybrid" adaptation of the multi-resolution Lausanne2008 atlas registered to MNI space \cite{hagmann2008mapping} in order to apply the hypergraph analysis. This 194 region ``hybrid" anatomical atlas minimizes variability in region size between subjects and brain regions \cite{davison_brain_2015}.

The functional data are preprocessed using FSL \cite{jenkinson_fsl_2012}, AFNI \cite{cox_afni:_1996} and Matlab \cite{MATLAB2015}. Preprocessing includes head motion correction with MCFLIRT \cite{jenkinson_improved_2002}, non-brain removal and spatial smoothing with AFNI 3dAutomask/3dDespike, slice-timing correction with AFNI 3dTshift, and additional motion artifact correction with AFNI 3dDetrend. 	
Additionally, each participant's time-averaged fMRI image is aligned to their structural T1 scan using FSL's FLIRT with boundary-based registration \cite{jenkinson_improved_2002, greve_accurate_2009}. The inverse of this transformation is applied to all participants' parcellation scales (generated in structural space) and parcellations are down-sampled into functional space with AFNI 3dfractionize. The mean signal across all voxels within a given brain region is calculated to produce a single representative time series. Time series for each task are concatenated to produce a single time series for each brain region.

\subsubsection*{Construction of Temporal Networks}
\begin{figure}[ht!]
	\centering
    \includegraphics[width=0.9\linewidth]{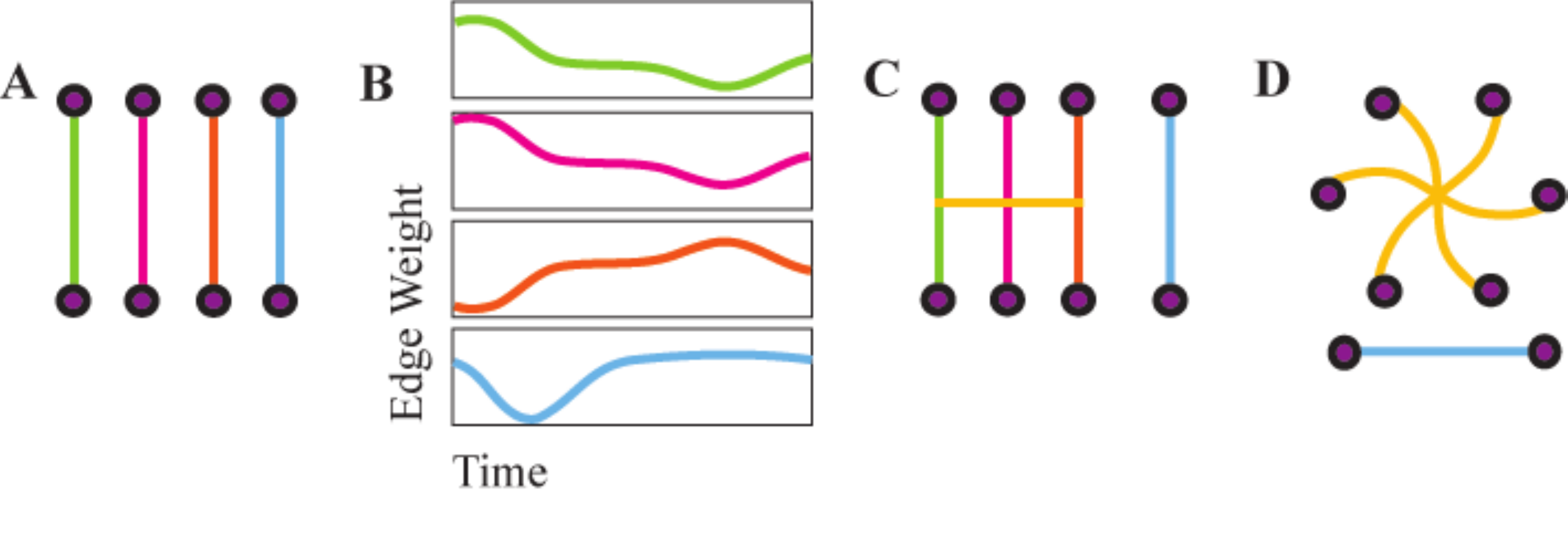}
	\caption{\textbf{Hypergraph construction:} Illustration of the method used to identify hyperedges. Edge weights are computed separately for each time window (A) and joined together to form edge weight time series (B). Significantly correlated edge time series are cross-linked to form a hyperedge (C). The group of hyperedges for an individual, with singletons removed, forms a hypergraph (D).}
	\label{fig1}
\end{figure}

For each subject, we construct a dynamic network model of brain function that accounts for changes in connectivity over time. Each of the $N = 194$ brain regions is a node in the network. The BOLD signal time series from each brain region is bandpass filtered to obtain data in the 0.06-0.125 Hz frequency range that contains task-related brain activity \cite{sun2004measuring,bassett2011dynamic,lynall2010functional,bassett2012altered}. This bandpassed time series is then windowed into one-minute sections. Node-node adjacency matrices of size $N\times N$ are constructed by taking Pearson's correlations between each pair of the $N = 194$ nodes in each of the time windows. Given the lengths of each scan, this windowing yields four rest, 18 attention, 18 word memory, and 18 face memory node-node adjacency matrices. The set of node-node adjacency matrices, one for each windowed section of time, represents the dynamic functional connectivity network; each edge, or pairwise connection between nodes, has an edge weight time series describing its temporal evolution across time windows, as depicted in Figure \ref{fig1}.

\subsubsection*{Hypergraph Construction}
Since hyperedges link edges in this dynamic network that have related temporal profiles, hypergraph structure is determined from the correlations between time-evolving weights of network edges \cite{bassett_cross-linked_2014} (See Figure \ref{fig1} for a schematic illustration of hypergraph construction). These are represented in an edge-edge adjacency matrix $\textbf{X}$, of size $E\times E$, where $E = N(N - 1)/2$ is the total number of possible edges in one time window of the DFC network. Each entry in $\textbf{X}$ is given by the Pearson correlation between the corresponding pair of edge weight time series in the DFC network. The $p$-values from these correlations are thresholded by a false discovery rate correction, which is more sensitive than other corrections for multiple comparisons and is thus effective for such neuroimaging network analyses \cite{genovese_thresholding_2002}. When the correlation between edges $i$ and $j$ is significant ($p <0.05$), we set $\xi_{ij} = X_{ij}$, to form the thresholded matrix $\xi$. All other elements of $\xi$ are set to zero. We binarize this thresholded matrix and obtain $\xi_{ij}^\prime$, where
\begin{align}
\xi_{ij}^\prime = \begin{cases}
1, & \text{if} \;\;\xi_{ij} \neq 0 ;\\ 0, & \text{if} \;\;\xi_{ij} = 0.
\end{cases}
\end{align}

Each connected component in the thresholded edge-edge correlation matrix $\xi^\prime$ -- that is, each set of edges with correlations between any two edges in the set but no significant correlation with edges in any other set -- forms a hyperedge. Taken together, all hyperedges in $\xi$ form a hypergraph. Since the edge weight time series are never thresholded and both high and low edge weights are preserved, hypergraphs provide information about edge dynamics without restricting the analysis to strong correlations in regional time series.

Our results are compared with a null model designed to ensure that hyperedges identified in our analysis can be attributed to system dynamics, rather than overall statistical properties of the data \cite{bassett_robust_2013}. To destroy temporal correspondences between edges but retain the mean and variance of each edge weight time series, the null model randomly reorders each edge time series individually and calculates correlations between the reordered edges.

Once hypergraphs are identified for each individual in the multi-task data set, hyperedges are classified according to whether the correlation in a cognitive state (i.e., rest or one of three cognitive tasks) is significant compared to a permutation null model over all states \cite{davison_brain_2015}. The hyperedges that satisfy these requirements are denoted as task-specific hyperedges, which we combine to form task-specific hypergraphs.

\subsubsection*{Hypergraph Metrics}
In this analysis, we examine several complementary measures on individual hypergraphs and focus on one method to extract meaningful information from the overall hyperedge distribution.

{\it Hyperedge size:} The size, $s(h)$, of a hyperedge $h$, is defined by
\begin{align}\label{eq:1}
s(h) = \sum_{i,j \in h} \xi^\prime_{i,j},
\end{align} where the sum is over the upper triangular elements of $\xi^\prime$, the binarized edge-edge adjacency matrix defined above. This is equivalent to the number of edges that are designated as part of this hyperedge.

{\it Singletons:} Singletons are hyperedges with $s(h) = 1$, edges with no significant correlation with any other edge in the network. We exclude singletons from the following analyses.

{\it Hypergraph cardinality:} The cardinality of an individual hypergraph is the number of non-singleton hyperedges present in the hypergraph.

{\it Hyperedge node degree:} The hyperedge degree of a node is the total number of hyperedges that contain that node.

{\it Task-specific hyperedges:} Hyperedges that exhibit a significantly higher correlation within one particular task are grouped into task-specific sets. The sets are calculated by using a permutation test to compare the correlation between edge time series for groups of edges in hypereges in a single task to the same correlation with edge time series data chosen randomly from all tasks. A Bonferroni correction for false positives due to multiple comparisons is employed to select task-specific hyperedges using the most stringent requirements \cite{hochberg_sharper_1988}.


\subsubsection*{Regression Procedure}
To investigate possible correlates of variability in individual hypergraph metrics, we perform a series of regression analyses. In each analysis, we use the hypergraph metric as the dependent variable and factors representing individual difference measures from the psychometric tests as the independent variables.

{\it Behavioral data categorization:} Behavioral and performance data for the multi-task study consist of 231 measures, while there are 115 measures for the age-memory study participants. There are 42 individual difference measures common to both studies, which we group into five categories, given in Table \ref{Table 1}. These categories are comprised of differing numbers of individual difference measures, which are summarized in Table \ref{Table_All}.

\begin{table}[ht!]
	\centering
	\caption{\textbf{Information retained for multi-task study:} Categories, number of factors for each, and how much overall variance from the multi-task individual difference data was retained for each category. Each category represents a subset of the 42 individual difference measures and the factors represent a percentage of the variance contained in the category for the multi-task data.}
	\begin{tabular}{|l|l|l|}
		\hline
		{\bf Category} & {\bf Factors} & {\bf Information Retained}\\
		\hline
		Performance & 2&  91.41\%\\
		Demographics & 2 & 92.62\%\\
		State of Mind & 3 & 80.45\%\\
		Cognitive Factors & 4 & 77.64\%\\
		Personality& 6 &  77.79\%\\
		\hline
	\end{tabular}
	\label{Table 1}
\end{table}

{\it Singular value decomposition:} Once the individual difference measures have been categorized, we demean all measures and perform a singular value decomposition (SVD) separately for each category.  We choose the minimum number of factors from the SVD for each category that retain at least 75\% of the variance across the category of measures from the multi-task study. Results from this process are presented in Table \ref{Table 1}.

{\it $R^2$ change:} The number of factors retained is not constant across categories, so we implement an adapted multivariate hierarchical regression \cite{raudenbush2002hierarchical,miller2012individual} to establish the comparative informativeness of each category. To assess the explanatory power of a given category, all factors in that category are held out for a ``control'' regression, and the difference in model $R^2$ between this reduced model and the full model is denoted as the contribution for that category. This corresponds to repeatedly performing a hierarchical regression with each category computed last, which gives a conservative estimate for the amount of variance attributable to the category \cite{miller2012individual}.

{\it Significance test:} To determine the significance of the regression coefficients, we use the $p$-values from $t$-tests on each multiple regression performed. The Bonferroni procedure for correcting for false positives due to multiple comparisons is used to adjust the $t$-test $p$-values over all regressions performed in this study \cite{hochberg_sharper_1988}. We employ the Bonferroni correction for multiple comparisons in all regression analyses because it is the most stringent test for significance.

\subsection*{Age-Memory Methods}

The majority of the methods are identical to those discussed for the multi-task data set. Below, we point out aspects that differ between the two analyses.

\subsubsection*{Age-Memory Experimental Design}

The word memory task in the age-memory study is constructed similarly to the word memory task in the multi-task data set. In addition to the memory task, participants completed a resting state scan and diffusion-tensor imaging, which we do not analyze further. Participants did not complete the face memory or attention tasks described in the first data set. The BOLD data were acquired while adult participants performed a recognition memory task with probabilistic cues. Prior to the scanning session, the participants studied 153 common English words, which were mixed with 153 novel lexical stimuli during the task. Participants were asked to determine whether the stimuli were studied or unstudied, with font color cues indicating whether the word had a 70\% probability or a 30\% probability of having been previously studied \cite{turner2015one}.

\subsubsection*{Image Acquisition and Processing}

Functional and structural data were collected from 126 healthy participants engaged in the word memory task. All functional data was acquired with a 3T Siemens TIM Trio MRI system with a 12-channel head coil. Scans consisted of T2*-weighted single shot gradient echo, echo-planar sequences sensitive to BOLD contrast (TR = 1.6 s; TE = 30 ms; FA = 90) with generalized autocalibrating partially parallel acquisitions (GRAPPA). In additon to the functional scans, high-resolution anatomical scans were performed for each participant using an MPRAGE sequence (TR = 2.3 s; TE = 2.98 ms; FA = 9; 160 slices; 1.1 mm thickness). Study participants also underwent behavioral assessments and psychological testing. Functional data from 31 participants were excluded due to technical issues, metal screening issues, claustrophobia, attrition, or lack of a complete individual differences survey. The results presented here are from 95 participants with usable functional and individual difference data. 

The functional data are preprocessed using FSL \cite{jenkinson_fsl_2012}, AFNI \cite{cox_afni:_1996}, and Matlab \cite{MATLAB2015}. Preprocessing includes head motion correction (MCFLIRT) \cite{jenkinson_improved_2002}, non-brain removal (BET) \cite{smith_BET_2002}, high-pass temporal filtering ($\sigma{} = 50\text{s}$), spatial smoothing, and grand mean intensity normalization (FEAT) \cite{wollrichBayesian2009}. Each voxel's time series is further denoised using a nuisance regression. The nuisance regression includes regressors for the six motion correction terms returned by MCFLIRT, their temporal derivatives, and the mean signal time series from the cerebrospinal fluid. The denoised data is registered to MNI space using FLIRT \cite{jenkinsonGlobal2001,jenkinsonOpt2002}. The T1 scan is first registered to the MNI template (12 df affine transformation), the functional data are registered with the T1 image (6 df affine transformation, trilinear interpolation), and the transformations are combined. As in the multi-task study, the mean BOLD signal across all voxels within a given brain region is calculated to produce a single representative time series. 

\subsubsection*{Construction of Temporal Networks}
Time series are demeaned and concatenated across the three functional runs of the word memory task to produce a single time series for each brain region. DFC networks are constructed here analogously to the multi-task study, with one key difference. In the age-memory analysis, we remove a single node-node adjacency matrix (i.e., a single time window) from the beginning and end of each functional run. This is to counteract edge effects from processing and ensure continuity across runs. We address this choice further in the \nameref{S1_Appendix} section of the Supporting Information.

\subsubsection*{Regression Procedure}
The regression procedure is similar to the analysis performed on the multi-task data. The individual difference data is kept in the common format, where only the 42 measures common to both studies are used and the categories are the same. Furthermore, the $R^2$ change and significance tests are calculated as above.

{\it Singular value decomposition:} We demean all measures and perform a singular value decomposition (SVD) on the combined multi-task and age-memory data separately for each category.  This differs from the multi-task analysis, where we only consider the variance retained over the multi-task data. We choose the minimum number of factors from each SVD that retain at least 75\% of the variance across both studies. Results from this process are presented in Table \ref{Table 3}.

\begin{table}[ht!]
	\centering
	\caption{\textbf{Factors common to the mutli-task and age-memory trials:} Categories, number of factors assigned to each, and how much of the overall variance was retained in each category. Each category represents a subset of the 42 individual difference measures and the factors represent a percentage of the variance contained in the category.}
	\begin{tabular}{|l|l|l|}
		\hline
		Category & Factors & Information Retained\\
		\hline
		Performance & 1 & 87.18\%\\
		Demographics & 1 & 86.14 \%\\
		State of Mind & 3 & 77.09\%\\
		Cognitive Factors & 3 & 81.25\%\\
		Personality& 4 &  78.56\%\\
		\hline
	\end{tabular}
	
	\label{Table 3}
\end{table}

\section*{Results}
As mentioned above, the hyperedge method has been applied to the multi-task data set in a previous study \cite{davison_brain_2015}. Here, we recapitulate the key findings from that investigation and provide results of exploratory analyses that motivate the followup analyses on the age-memory data set. We then present results from the age-memory analysis.
\subsection*{Summary of Prior Results}

A previous study of the multi-task data identified measures that capture significant differences in population-level hypergraph structure across tasks \cite{davison_brain_2015}. Furthermore, extensive variation was observed in several hypergraph measures, including hypergraph cardinality, across individuals. These results emphasize that hypergraph structure can be used to differentiate between task states and motivates our investigation of the correspondence between hypergraph structure and individual difference measures.

\begin{figure}[ht!]
	\centering
	\includegraphics[width=0.9\linewidth]{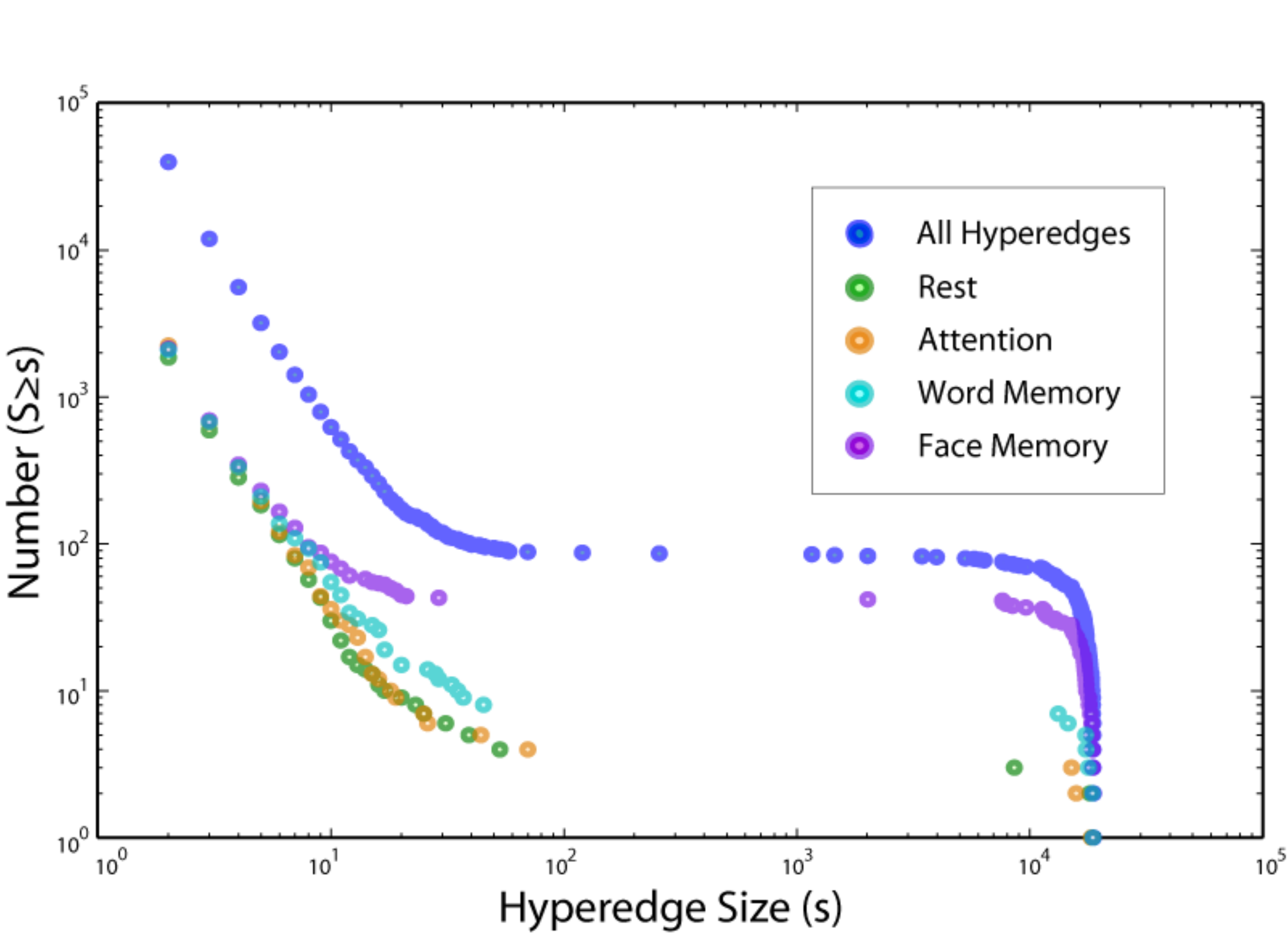}
	\caption{\textbf{Multi-task cumulative size distribution:} The empirical cumulative distribution function of hyperedge sizes for all subjects in the multi-task study. Also shown is a trace for the empirical cumulative distribution functions of hyperedge sizes over all subjects for each of the four task-specific hypergraphs.}
	\label{fig2}
\end{figure}

Figure \ref{fig2} depicts the empirical cumulative hyperedge size distributions for all hyperedges found across all subjects in the multi-task data set. As a null test, we shuffle the data over time and find no hyperedges of size greater than one. There is a rough power law for the smaller sizes ($s < 100$), followed by a gap in the distribution from about 100 to 1000 and a sharp drop at the system size ($s = \binom{194}{2} = 18721$). The shape of the distribution is due to the consistent hypergraph structure across individuals; the majority of subjects in this study have a hypergraph composed of one large hyperedge and many small hyperedges. While this characteristic structure is common to most subjects in the study, the size of the largest hyperedge varies across individuals. This size is closely related to the hypergraph cardinality, defined as the number of hyperedges in a hypergraph, a measure which also exhibits large variation.

Figure \ref{fig2} also depicts task-dependent differences in the cumulative size distributions of task-specific hyperedges. Memory-specific hyperedges tend to be more numerous than those specific to the rest and attention tasks. However, the total number of task-specific hyperedges for any task is at least ten times fewer than the total number of hyperedges. Our strict definition of task specificity includes only hyperedges specific to a single task and discards those associated with more than one task. This approach is conservative, and likely leaves some meaningfully task-related hyperedges unclassified. However, it reduces the complexity of the task-specific results, and provides greater confidence that any hyperedges classified as task-specific are indeed providing truly task-driven information due to coherence within that task alone, rather than coherence due to an unrelated driver that is common to several tasks.

\begin{figure}[ht!]
	\centering
    \includegraphics[width=0.9\linewidth]{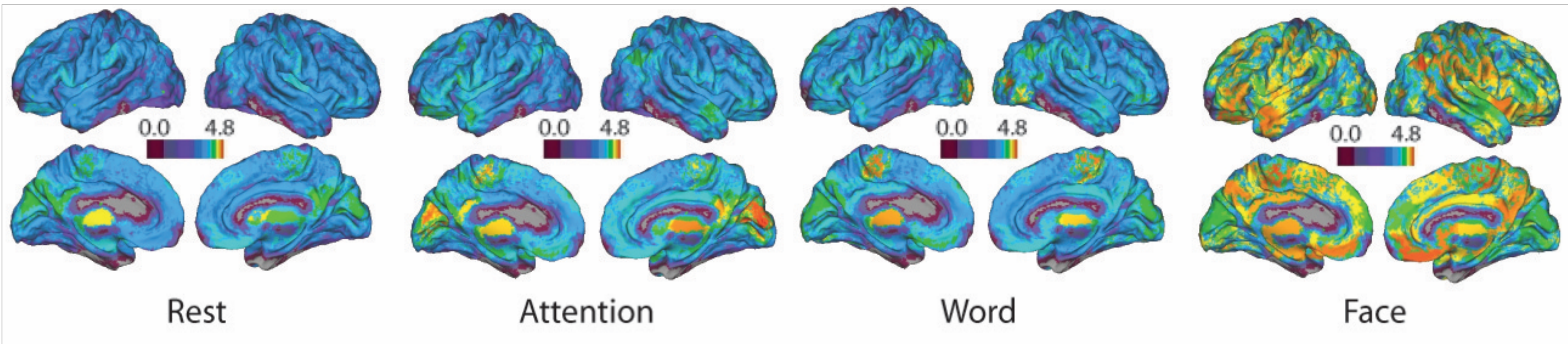}
	\caption{\textbf{Node degree spatial distribution:}  Here, the number of hyperedges at each node over all individuals in the multi-task study is plotted on the brain. The scale is logarithmic, and higher values in a region indicate that there are more hyperedges that include the region.}
	\label{fig3}
\end{figure}

There are significant differences in the spatial organization of task-specific hyperedges over all individuals that are visualized in Figure \ref{fig3}. The plots depict task-specific hyperedge degree across the brain for each of the four tasks. In addition to the differences in magnitude between word memory and the other tasks, the locations of high hyperedge concentration vary with task.

These significant differences in hypergraph structure between the tasks confirm that hypergraph structure varies between task states. However, persistent variability in hypergraph measures across individuals indicates that the hypergraph method reflects innate differences beyond the current task state. The work presented here follows this line of inquiry, beginning with an analysis of individual differences in the multi-task data set.

\subsection*{Multi-Task Results: Individual Differences}

Here, we illustrate and quantify the wide variation in hypergraph measures across individuals in the multi-task data. In brief, we identify a particular measure, hypergraph cardinality, that demonstrates large variance across all individuals but is consistent within individuals. Following this, we investigate relationships between the variation in individual difference measures and the variation in hypergraph cardinality. The results from this study are not statistically significant due to the limited variation in individual difference measures and strict corrections for multiple comparisons. However, we report a marginally significant result relating demographics and word-memory hyperedge cardinality that motivates further analyses on the age-memory data set.

\subsubsection*{Individual Variability and Consistency in Hypergraph Metrics}

\begin{figure}[ht!]
	\centering
    \includegraphics[width=0.9\linewidth]{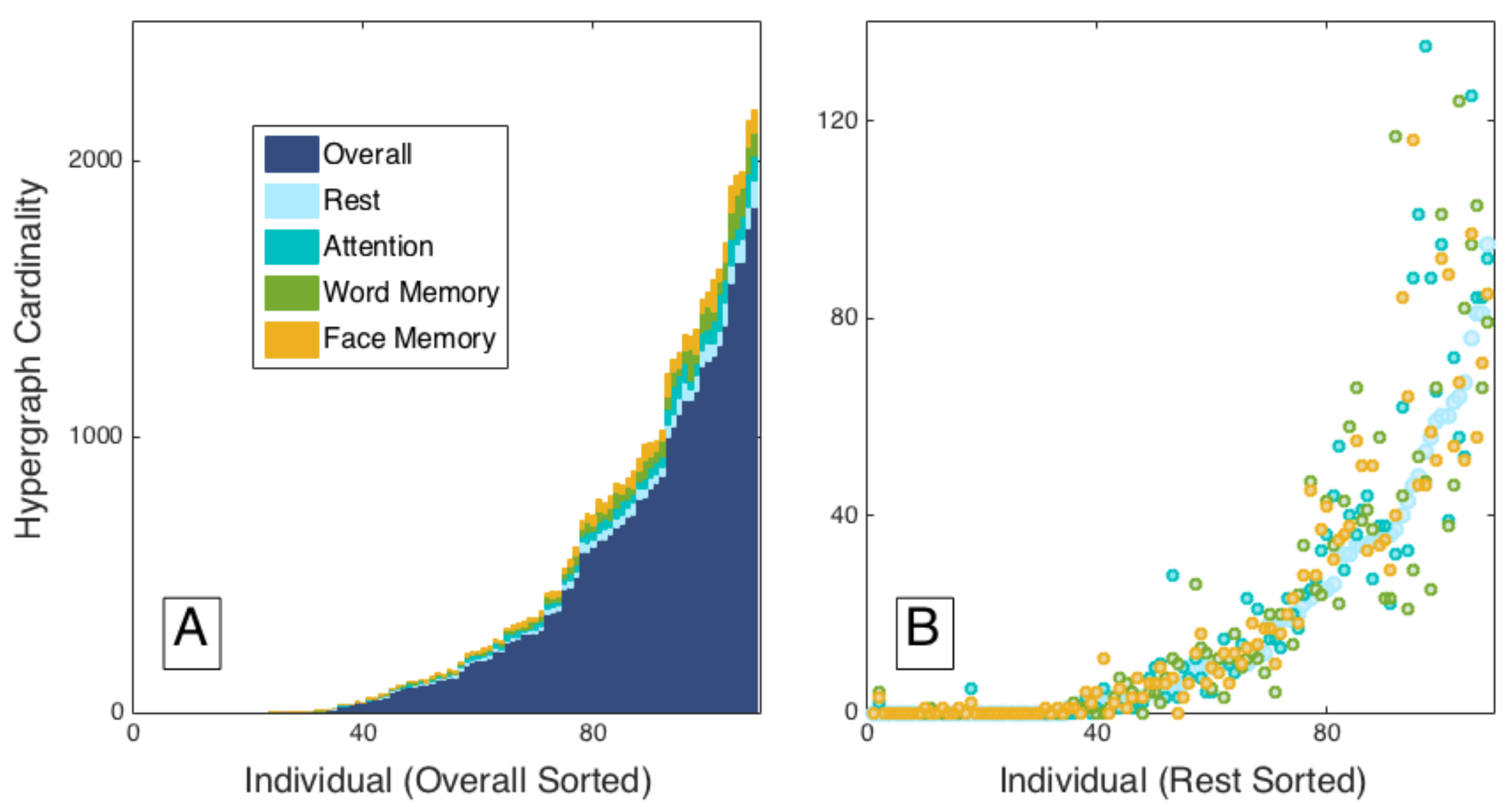}
	\caption{\textbf{Individual variability:} Hypergraph cardinality for overall hyperedges (A), sorted from smallest to largest cardinality. The plot also includes task-specific cardinalities sorted by overall cardinality. Panel (B) depicts the cardinality for task-specific hyperedges, sorted by rest cardinality. The number of hyperedges across tasks is fairly consistent within individuals, in contrast to the range of hyperedge number across individuals.}
	\label{fig4}
\end{figure}

Although our previous study focused on group-level properties of hypergraphs across tasks, notable individual differences in functional dynamics were also seen \cite{davison_brain_2015}. Here, we confirm those preliminary observations by investigating the hypergraph cardinality measure and finding that it displays extreme variations across subjects in the multi-task data set, as shown in panel (A) of Figure \ref{fig4}. These individual variations in hypergraph cardinality span several orders of magnitude. 

Despite this large variation between participants, hypergraph cardinality follows a consistent pattern within each participant across tasks. Panel (B) of Figure \ref{fig4} depicts individual measures of hypergraph cardinality for hyperedges specific to each task, with subjects sorted by rest hypergraph cardinality. Within participants, the task-specific hypergraph cardinality is consistent across task states and follows the distribution for rest-specific hyperedges, which further emphasizes the consistency of hypergraph cardinality within individuals.

Consistent hypergraph cardinality within participants over all tasks indicates that there are characteristics specific to individuals that drive hypergraph properties, even in designated task-specific hypergraphs. These patterns imply the existence of driving influences on hypergraph structure that are independent of performance on a specific task. To investigate this further, we examine how individual difference measures from demographic and behavioral data relate to hypergraph cardinality.

\subsubsection*{Drivers of Individual Variability}
To investigate possible sources of the large variation in hypergraph cardinality seen above, as well as to quantify the extent of the consistency of hyperedge cardinality across tasks, we perform a series of multiple regression analyses on the multi-task data, as described in Methods. 

First, using the cardinality of task-specific hypergraphs as the dependent variable, we perform a regression analysis for each non-resting task (attention, word memory, and face memory) that includes the cardinality of the rest-specific hypergraph and the factors shown in Table \ref{Table 1} as independent variables. Table \ref{Table 2} gives the $R^2$ change values and $p$-values associated with the rest predictor for each task-specific regression. In all three tasks, the rest predictor alone significantly explains the variance in task-specific hypergraph cardinality. This confirms and quantifies our observation in Figure \ref{fig4} that hypergraph cardinality is consistent across each individual's task-specific hypergraphs---i.e., it is trait-like. The individual difference measures used as independent variables are not significant after the Bonferroni correction for multiple comparisons over all tests. However, including the rest-specific hypergraph cardinality, which is closely linked to overall hypergraph cardinality, as an independent variable in the regression accounts for the variation across individuals that is consistent across tasks.

\begin{table}[ht!]
	\centering
    \includegraphics[width=0.9\linewidth]{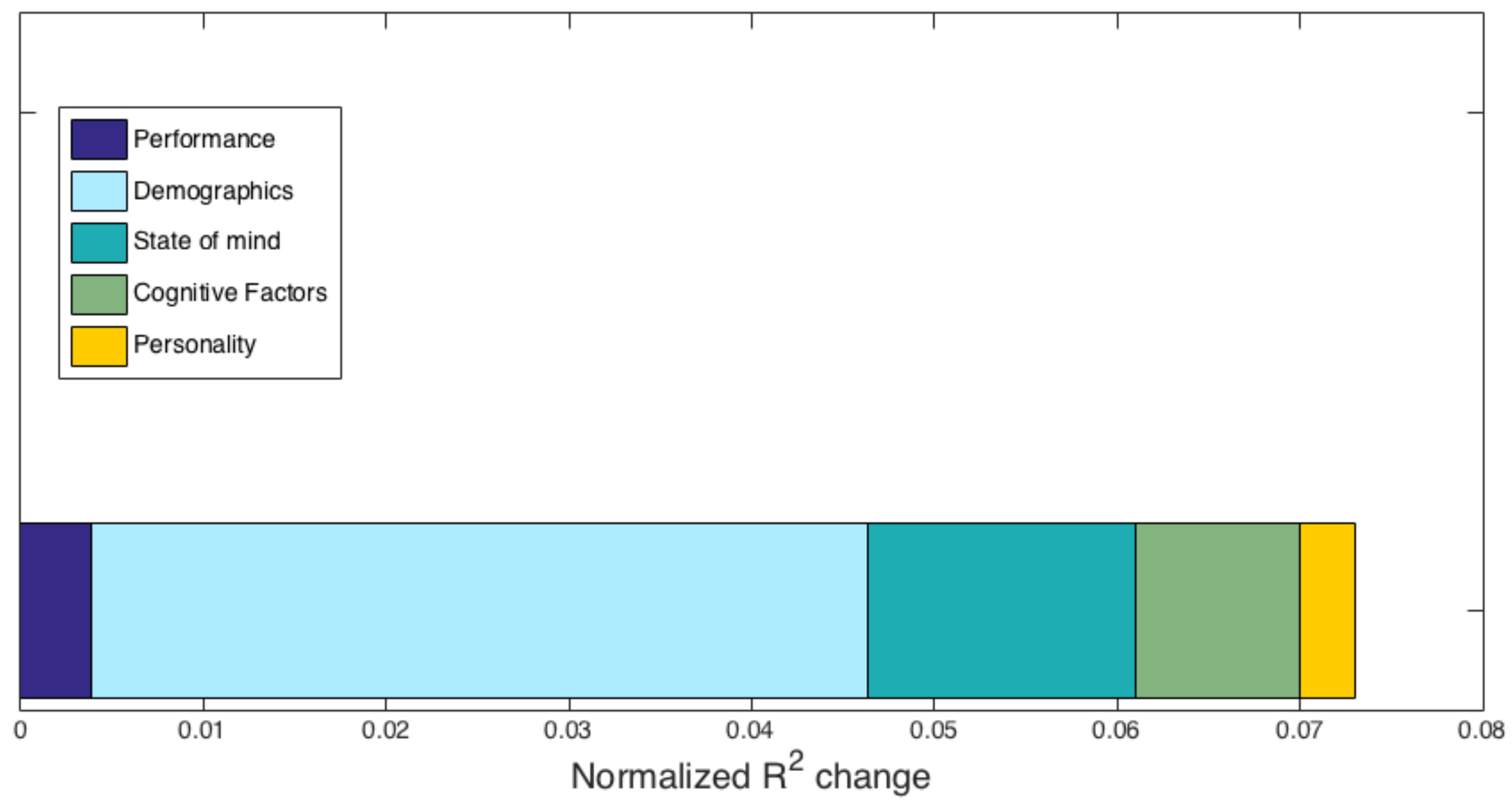}
	\caption{\textbf{Rest regression $R^2$ values:} $R^2$ values for the regression between rest-specific hyperedge cardinality and hyperedge cardinality for each of the other three tasks.}
	\begin{tabular}{|l|l|l|l|}
		\hline
		& {\bf Attention} & {\bf Word Memory} & {\bf Face Memory}\\
		\hline
		{\bf $R^2$ change} & 0.72 & 0.58 & 0.68\\
		\hline
		{\bf $p$-value} & $<0.0001$ & $<0.0001$& $<0.0001$ \\
		\hline
	\end{tabular}
	
	\label{Table 2}
\end{table}

To identify possible drivers of this individual variation, we perform another regression analysis, using the individual difference measures from Table \ref{Table 1} as independent variables and overall hypergraph cardinality as the dependent variable. Figure \ref{fig5} depicts the $R^2$ changes from this analysis for each category of factors. The $t$-test identifies no factors with significant correspondence to hypergraph cardinality, but we observe that the demographics category has the largest $R^2$ change. The $t$-test $p$-value for one of the factors in the demographics category is $<0.05$ and is by far the lowest $p$-value in this stage of the analysis. However, due to our stringent requirements for correcting for multiple comparisons and the number of tests we performed, this correlation is not statistically significant. The marginally significant demographics factor has a loading of $-0.95$ for the age measure and $	-0.31$ for the years of education measure; the loading for sex and handedness demographic measures are comparatively negligible, with magnitudes $< 0.02$.

\begin{figure}[h!]
	\centering
	\caption{\textbf{Multi-task $R^2$ changes:} Normalized $R^2$ changes with respect to hypergraph cardinality across individuals in the multi-task study. $R^2$ changes are calculated from the regression procedure outlined in Methods, with five distinct categories common to the multi-task and age-memory studies. The largest normalized $R^2$ change is from the demographics factor.}
	\label{fig5}
\end{figure}

\subsubsection*{Summary of Multi-task Results}

On the basis of our previous results applying hyperedge analysis to this data set, which hints at substantial variability across individuals in hypergraph structure (Figure \ref{fig2}), we carry out several regression analyses designed to identify individual drivers of this variability. There were two key results. The first result is that overall and task-specific hypergraph cardinality show notable variation between subjects, but remarkable consistency within subjects for all tasks (Figure \ref{fig4}).

The second key result from this exploratory analysis is the finding of a marginally significant relationship between the demographics category and hyperedge cardinality. Limits to the explanatory power of the multi-task data set may be determined by limited variation in some demographic measures -- particularly the small range (27--45) and variance (19) in subject age, which poorly represents the ages observed in the entire population. We thus extend our analysis to a complementary data set collected on a longer study of the word memory task with participants aged 18--75. In the next section, we report the results of our independent analysis of this age-memory data set, which confirm the relationship between age and hypergraph cardinality suggested by the multi-task results.

\subsection*{Age-Memory Results}

To supplement the findings from the multi-task data set, we perform a parallel set of analyses on the age-memory data set. The data set includes participants with ages ranging from 18 to 75, a range three times larger than the range of ages in the multi-task study. Furthermore, the age-memory study uses an almost identical task to the multi-task word-memory task. In this section, we combine hypergraph results for all participants in the age-memory data set and obtain a distribution of hyperedge size over all participants with similar features to the hyperedge size distribution from the word-memory task of the multi-task data. We then identify and test specific drivers of individual variation in hypergraph cardinality for the age-memory study participants. We find a strong correspondence between age and hypergraph cardinality that confirms the preliminary result from the multi-task study.

\subsubsection*{Hypergraph Statistics}

The cumulative size distribution of hyperedges for all individuals in the age-memory study is depicted in blue in Panel (A) of Figure \ref{fig6}. To compare these age-memory hyperedges with the word memory portion of the multi-task study, we identify a new set of hyperedges using only the portion of the multi-task functional time series recorded during the word-memory task for each subject; the distribution of sizes for these hyperedges are plotted in pink. Note that these new word-memory hyperedges from the multi-task data are fundamentally different from the ``word memory-specific'' hyperedges depicted in Figure \ref{fig2}. The ``word memory-specific'' hyperedges are those hyperedges computed over all tasks, but classified to be driven by correlations in the word memory task alone. In contrast, the new word-memory hyperedges in Figure \ref{fig6} are found by using just the word-memory subset of the multi-task data, with no further classification applied.

The distributions of sizes are similar at smaller size scales, but differ somewhat at larger size scales. There are many more hyperedges close to the system size in the age-memory task, while the word-memory hyperedges from the multi-task data set tend to be smaller. The length of the multi-task word-memory time series is shorter than the age-memory time series, which may contribute to this effect \cite{birn2013effect}. To investigate the size distributions without the effect of full-brain hyperedges, we remove the largest hyperedge from each subject's hypergraph and plot the resulting distribution in Panel B of Figure \ref{fig6}. With this adjustment, the distribution of age-memory hyperedge sizes has a striking agreement with the size distribution of hyperedges constructed from the multi-task word memory data. In both distributions, there is power law behavior for small sizes, similar to that observed in Figure \ref{fig2}. Furthermore, the distributions without the largest hyperedges are almost identical; the power of the fit to multi-task word memory data is $-2.21$ and the intercept is $7.91\times 10^{4}$, while the power of the fit to the age-memory data is $-2.37$ and the intercept is $1.46\times 10^{5}$.

We construct a null model, as detailed in the multi-task Methods section, by temporally shuffling the data and find no hyperedges with size greater than one, indicating that the hyperedges identified in the unshuffled data are capturing statistically significant aspects of brain dynamics. In addition, the close correspondence between these two distributions of word-memory hyperedges suggests that the analysis captures aspects of brain dynamics that are robust across imaging sessions and populations.

\begin{figure}[h!]
	\centering
	\caption{\textbf{Comparison of cumulative size distribution:} Panel (A) depicts the cumulative distribution of hyperedge sizes over all individuals in the age-memory study compared with the sizes of the set of hyperedges constructed from only the word-memory task of the multi-task data set. Panel (B) illustrates the cumulative distribution of sizes for all individuals in both studies with the largest hyperedge for each individual subject removed. When this is done, the distributions overlap and are well described by a power law with close alignment in slope and magnitude across studies.}
	\label{fig6}
\end{figure}

\begin{figure}[ht!]
	\centering    
    \includegraphics[width=0.9\linewidth]{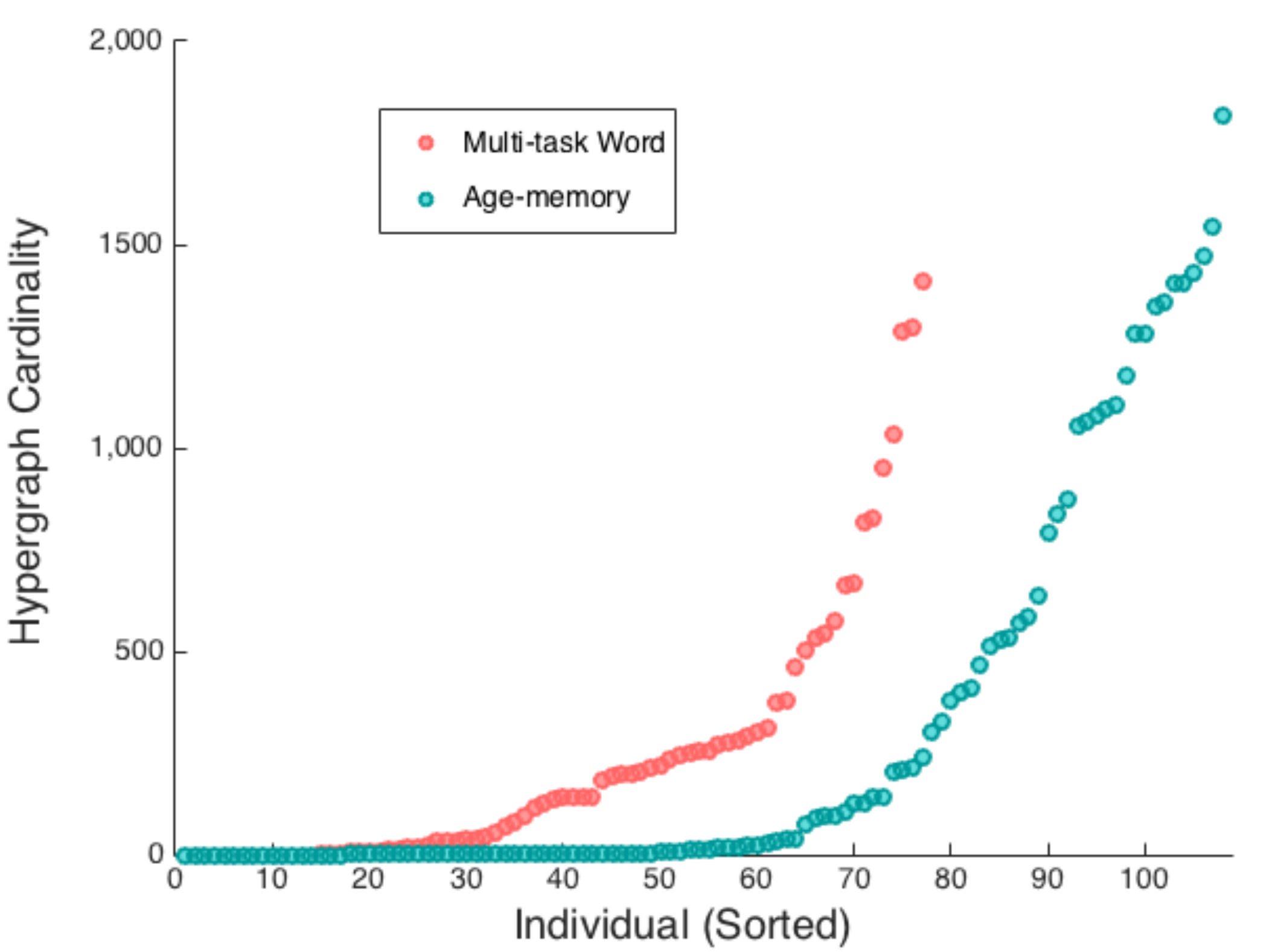}
	\caption{\textbf{Sorted hypergraph cardinality:} Increasing hyperedge cardinality for individual multi-task word-only and age-memory hypergraphs. The variability for both studies is similar to the variability in multi-task overall hypergraph cardinality, depicted in Panel (A) of Figure \ref{fig4}.}
	\label{fig7}
\end{figure}

The inter-subject variability in multi-task hypergraph cardinality spanned several orders of magnitude and followed consistent patterns within subjects for differing cognitive states. We compare the individual hypergraph cardinality for the age-memory and multi-task word-only studies in Figure \ref{fig7}. In the age-memory data, hypergraph cardinality ranges from 0 to 1817, which is a similar range of variability as that observed for the complete overall multi-task data set in Figure \ref{fig4}. There are 79 subjects with nonzero hyperedge cardinality, indicating that significant non-singleton hyperedges are present in less than two thirds of the subjects. For the remaining analyses, we only consider the 79 subjects with nonzero hypergraph cardinality. For the overall hypergraphs, hypergraph cardinality ranges from 0 to 1832. The maximum hypergraph cardinality for the multi-task word-only data is 1408, which is markedly less than that observed for the age-memory data and may be a result of the shorter time series for the multi-task word task. The presence of near-system size hyperedges, which may also be due to the shorter multi-task word time series, affects hypergraph cardinality by resulting in hypergraphs with cardinality near one.

\subsubsection*{Age-Memory Hypergraph Correspondence With Age}

Having confirmed that hypergraph composition is similar for the multi-task word study and the age-memory study, we investigate whether the individual variability in hypergraph cardinality seen in Figure \ref{fig7} corresponds to individual difference factors for the age-memory study.

We perform a multiple regression on the 12 factors distributed across five categories in Table \ref{Table 3}. Head motion has been found to induce correlations in FC analyses \cite{power2012spurious}, and a previous study using this data found a significant correlation between age and amount of head motion during the experiment \cite{turner2015one}. To ensure that excessive head motion is not contributing to our result in any way, we include head motion (operationalized as the average relative movement as computed by MCFLIRT) as a predictor in this regression.

\begin{figure}[ht!]
	\centering
    \includegraphics[width=0.9\linewidth]{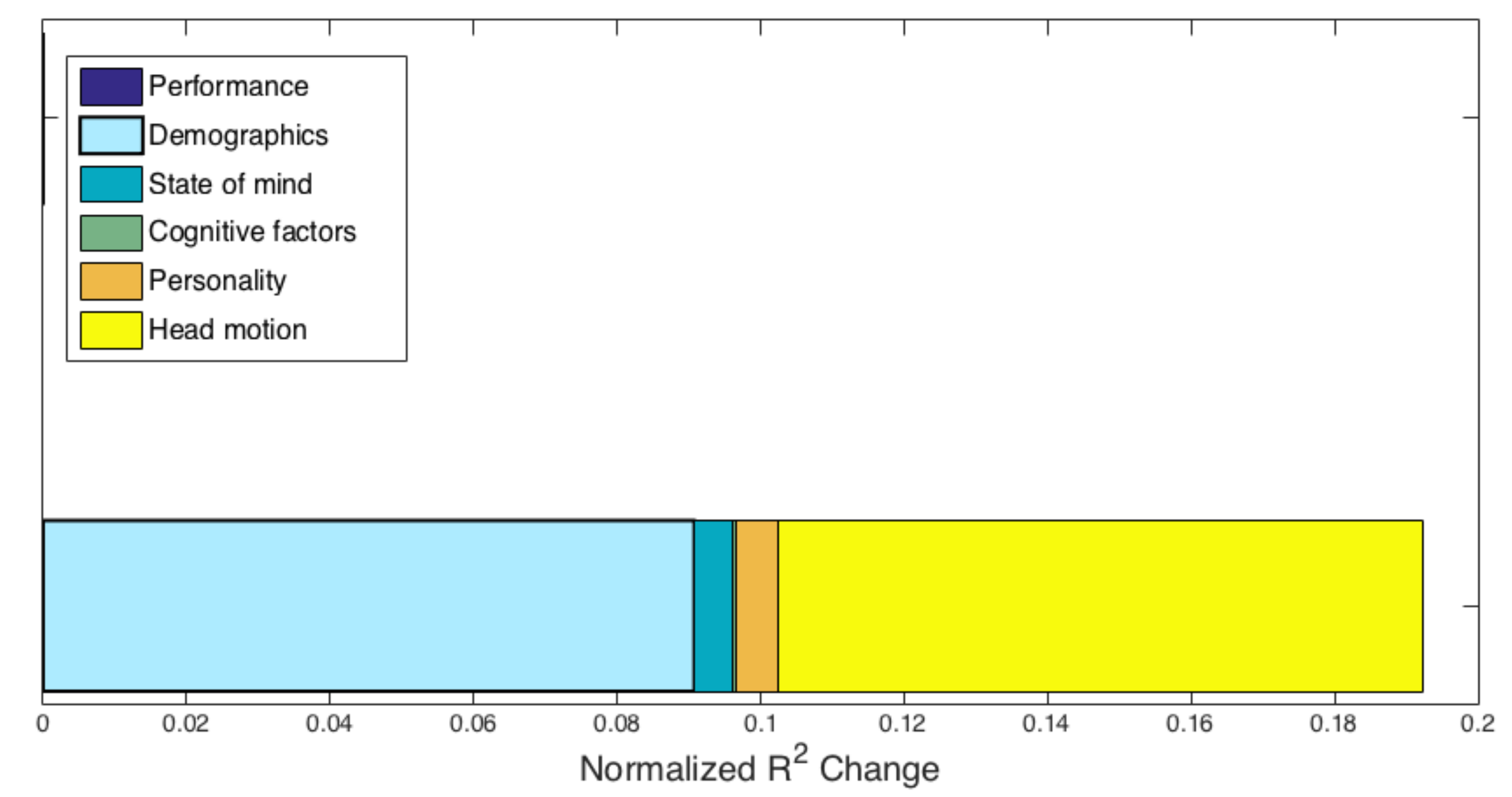}
    \caption{\textbf{Age-memory $R^2$ changes:} Normalized $R^2$ changes with respect to hypergraph cardinality across individuals in the age-memory study. The largest normalized $R^2$ changes are from the demographics factor and head motion measure. The demographics factor is the only significant predictor of hypergraph cardinality, which we denote with a bold outline.}
	\label{fig8}
\end{figure}

The overall $R^2$ value for the multiple regression analysis was 0.3452, indicating that the predictors explain about a third of the variance in the overall data. After a Bonferroni correction for multiple comparisons across all regression studies included in this paper \cite{hochberg_sharper_1988}, the demographics factor is the only significant predictor of hyperedge cardinality. The normalized $R^2$ changes for hypergraph cardinality can be seen in Figure \ref{fig8}; the demographics factor has the largest normalized $R^2$ change and the only significant $p$-value ($<0.005$) in the regression. These results correspond with the marginal result from the multi-task data set, where the demographics factor is a marginally significant predictor.

Much of the variation in the demographics factor (73.5\%) is directly attributable to age. We attempt to isolate the specific relationship between age and hypergraph cardinality by performing a separate regression. In this regression, hypergraph cardinality is the dependent variable and the independent variables are age and head motion. The relationship between age and hypergraph cardinality is significant, with the $t$-test $p$-value well below the Bonferroni correction over all regression analyses presented in this work, at $p <0.001$.

\begin{figure}[ht!]
	\centering
    \includegraphics[width=0.9\linewidth]{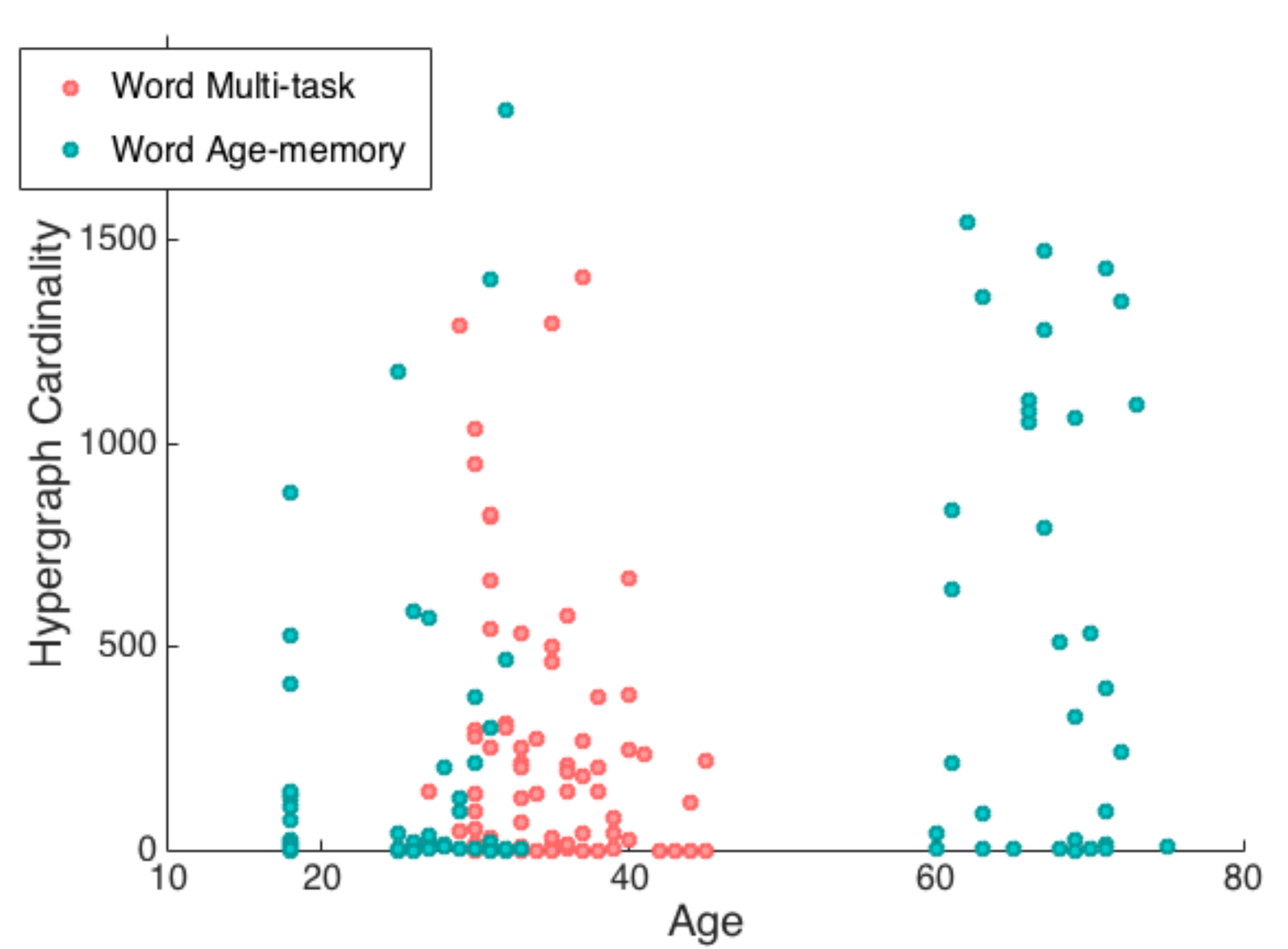}
    \caption{\textbf{Hypergraph cardinality and age:} Scatter plot of hypergraph cardinality as a function of age for the age-memory data set and word memory task of the multi-task data set. The correspondence between increasing age and larger hypergraph cardinality can be observed.}
	\label{fig9}
\end{figure}

This is a positive relationship, indicating that older individuals tend to have higher hypergraph cardinality, while younger participants tend towards lower hypergraph cardinality. An illustration of this correspondence between hypergraph cardinality and age is presented in Figure \ref{fig9}. As age increases, the number of hyperedges in a participant's hypergraph increases as well.  We verify that this relationship holds beyond this particular study by reintroducing the word-memory data from the multi-task study and performing a correlation between hypergraph cardinality and age over both studies. Age and hypergraph cardinality have a Spearman correlation coefficient of $\rho = 0.32$, and the $p$-value for this correlation, $ p <10^{-5}$, is significant when we use the Bonferroni correction over all analyses presented in this paper.

\section*{Discussion}

Improving our understanding of the drivers of individual differences in functional brain imaging data can give insight into the mechanisms that lead to individual behavior. Dynamic FC has been used over groups to explain changes in the brain attributed to individual differences in learning \cite{bassett2011dynamic, bassett2013task, mantzaris2013dynamic}. Hypergraphs in particular have been used to analyze how long-term learning impacts the functional network structure \cite{bassett2013task} and how the brain switches between cognitive states \cite{davison_brain_2015}. Here, we extend previous dynamic FC studies that showed common properties of the dynamics at the level of the group \cite{davison_brain_2015} to investigate the drivers of strong individual variations in certain hypergraph metrics.

\subsection*{Disparate Sources of Variability in Hypergraph Structure}

As we showed in the Multi-Task Analysis, the hypergraph cardinality varies widely across individuals, but is consistent between task states. Previous work on the multi-task data set found that the probability for hypergraphs to appear in a particular network configuration over individuals was significantly different depending on task state \cite{davison_brain_2015}. Consistent spatial organization rules for each task existed at the level of the group. There were similarities in the spatial arrangement of hyperedges in the brain for differing tasks, but certain properties were found to vary significantly between tasks. Brain areas in the occipital lobe in particular were highly likely to participate in the hypergraph network across individuals and across tasks, likely due to the visual nature of most of the cognitive tasks studied.

Here, we study hypergraph cardinality, which displays high variability across individuals and consistency across tasks within individuals (Figure \ref{fig4}). This indicates that hypergraph cardinality serves as an individual signature of a subject's brain dynamics. The similarities across subjects in the spatial distributions of hypergraphs described in \cite{davison_brain_2015} capture information orthogonal to the information summarized by hypergraph cardinality. For example, there are some individuals for whom the visual brain regions are linked by many hyperedges, and some for whom those same regions are linked by relatively few hyperedges, but these regions are more likely than others to be included in hypergraphs in the majority of subjects. This suggests that, for some subjects, brain regions tend to be more dynamically integrated in general, with co-varying functional relationships across many brain circuits; in other subjects, connectivity dynamics are more fragmented across the brain.

The high degree of variability in hypergraph cardinality across subjects and consistency within subjects, combined with the significant differences in spatial hyperedge arrangement across tasks, indicate that hypergraphs are a useful analysis tool for investigating both individual and task-based differences in brain function in a variety of settings. At the same time, hypergraphs can provide a view of dynamic patterns that complements other commonly used DFC methods. For example, many FC methods exclusively investigate the structure of strong correlations in functional data \cite{hermundstad2014structurally, bassett_robust_2013, greicius2009resting, bassett2008hierarchical}; hypergraph analysis captures information about both strongly and weakly correlated dynamics and how sets of brain regions transition between them \cite{bassett_cross-linked_2014}.

Although they are highly informative, many of the hypergraph metrics we study here are representative measures that greatly reduce the complexity of the hypergraph and only reveal a small part of the information contained in its structure. Further development of methods to utilize more of the information that hypergraphs provide will allow characterization of the consistency of particular hyperedges and dynamic modes, an understanding of which are important for behavior, or influenced by demographics or disease. Future work is also needed to further quantify the spatial differences in hypergraph arrangement across both individuals and tasks, to clarify the extent of overlap between the two types of information, and to determine whether the individual variability in cardinality can be mapped to individual spatial differences in hypergraph structure.

\subsection*{Relationship Between Age and Changes in DFC Networks}
FC studies have established clear trends associated with aging, including a decrease in connectivity within functional networks and an increase in connectivity across different functional networks in resting and task states \cite{geerligs_reduced_2014, madden_adult_2010, meunier_age-related_2009, ng_reduced_2016, betzel_changes_2014}. Many of these studies have considered resting-state FC, because the absence of task stimulus provides a simple and reliable setting for comparison between subjects \cite{ferreira_resting-state_2013}, although recent studies have successfully used FC networks to study various cognitive proceses \cite{medaglia_cognitive_2015}. The default mode network (DMN) and similar resting-state analyses may miss functional changes evoked by task states; while the DMN FC decreases with age, task-related sensorimotor network FC has been shown to increase with age \cite{song2014age, geerligs2015brain}. Similarly, FC in memory tasks shows increased segmentation with age \cite{sala-llonch_changes_2014}. Extending these analyses to incorporate the dynamics of functional interactions is a necessary step towards quantifying individual changes in functional brain dynamics associated with age.

Several efforts have been made to capture individual age-related differences with methods from dynamic FC. Dynamic community structure and amplitude of low-frequency fluctuation of FC were both found to be strongly correlated with age, illustrating that functional dynamics are closely linked with aging \cite{qin2015predicting, Schlesinger2016age}. In the dynamic community detection analysis, functional communities were found to be more fragmented with age, which agrees with the hypergraph cardinality result presented here \cite{Schlesinger2016age}.  A multi-scale community detection analysis uncovered similar fragmentation with age for small scales \cite{betzel2015functional}. Our finding that hypergraph cardinality also increases with age aligns with this result and provides further information based upon its ability to capture higher-order dynamic patterns across larger ensembles of brain regions. Not only do the functional similarities of communities of brain regions themselves become less distinct as humans age, but the temporal profiles of these functional similarities also become less integrated across brain regions. The agreement of this result with known age-related changes in FC \cite{dennis2014functional, contreras2015structural, sala2015reorganization, ferreira2015aging,ng_reduced_2016} demonstrates the ability of hypergraph methods to capture and quantify major brain changes. Moreover, since the hypergraph analysis is not limited to strong correlations, our analysis further suggests that age is related not only to the organization of functional activity in groups of brain regions with strongly coherent activity, but also to the coordination between groups of regions that transition from being strongly to weakly correlated over time (or {\it vice versa}).

The reported correspondence between age and hypergraph cardinality is significant in the age-memory data set, but our analysis did not include data that could verify this relationship for cognitive tasks other than the word memory task. Although memory is a cognitive ability known to decline with age in many individuals, it is unlikely that the specific task studied in the age-memory data set drives this result. Rather, the consistency of hypergraph cardinality across tasks seen in the multi-task data set in Figure \ref{fig4}(B) suggests that similar hypergraph cardinalities may be found during other tasks in data sets with higher age variability, and that the relationship between age and cardinality is unlikely to depend primarily on the behavioral task. Further investigation is needed to determine whether individual differences in hyperedge structure have any significant relationship to behavioral or cognitive performance on any particular task.

\section*{Conclusion}
Here, we have shown that the considerable differences in functional connectivity dynamics across individuals are closely linked with age. The hypergraph method is presented as a complex analysis tool that captures information about group-level similarities that differ between task states as well as individual differences that are consistent within individuals, across tasks. Further investigation into a single hypergraph metric (hypergraph cardinality) that varies across individuals uncovers a significant relationship between hypergraph cardinality and age. Specifically, there are a greater number of hyperedges in older individuals' hypergraphs, suggesting that there are more small groups of regions with cohesively evolving dynamics and indicating a loss of coherence across larger, spatially distributed intrinsic functional connectivity networks. This complements widely reported relationships between FC and human aging by providing new insight into how FC activity and the co-evolution of FC activity are altered with increasing age, including the loss of large groups of co-evolving brain regions in older individuals. The correspondence with and extension of classic FC results to new dynamic regimes, along with the unique capacity of hypergraphs to probe multiple dimensions of both strong and weak dynamic variability, show that hypergraph analysis is a valuable tool for understanding age-related changes and other individual differences in dynamic brain function.

\section*{Supporting Information}
\beginsupplement

\begin{figure}[ht!]
	\centering
    \includegraphics[width=0.9\linewidth]{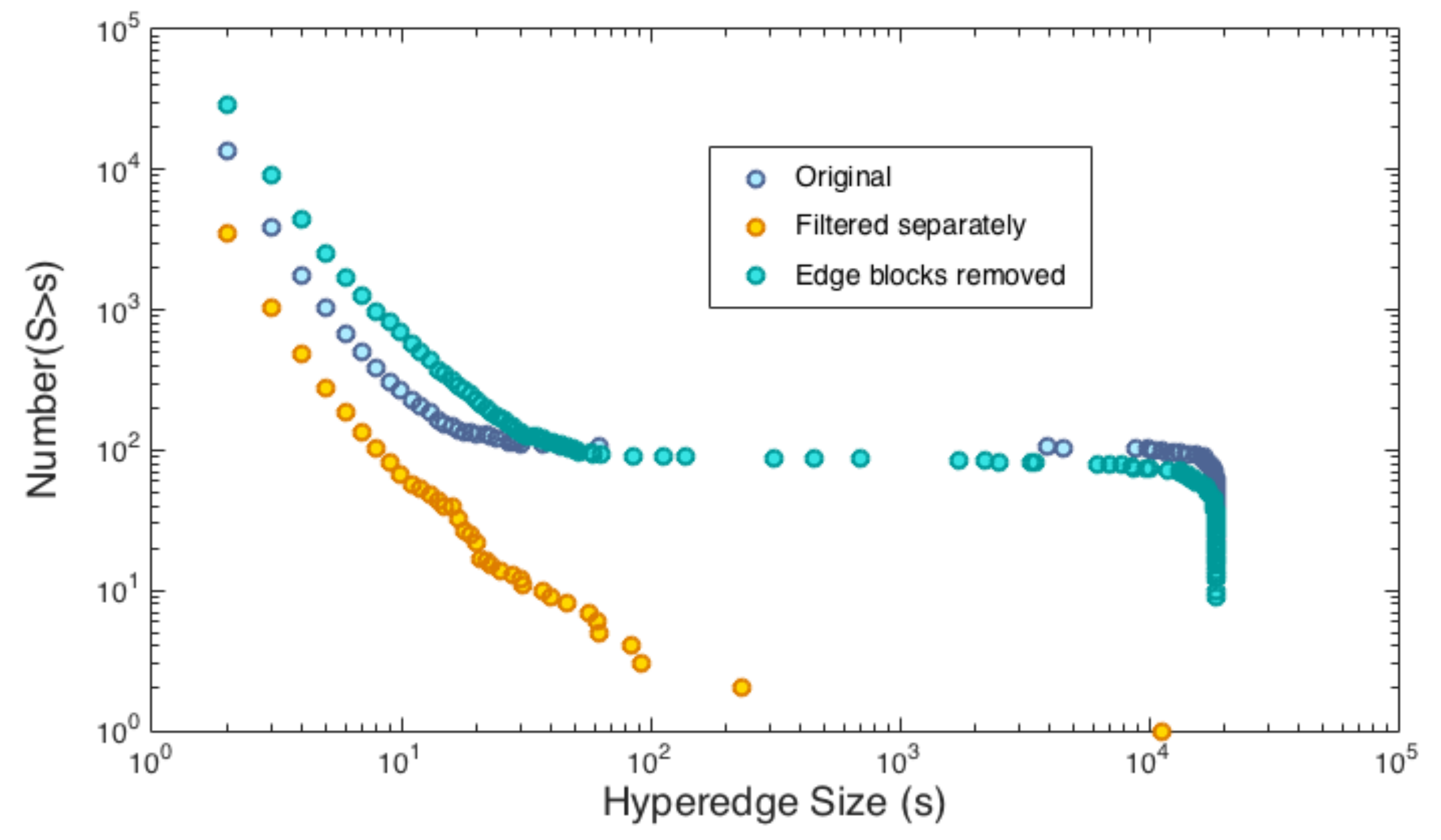}
    \caption{\textbf{Edge compensation comparison:}  Cumulative size distributions for the original age-memory data set (with no changes to remove effects of the edges) and two methods for removing potential effects from the edges. The ``edge blocks removed" method is used in all analyses in the main text.}
	\label{figS1}
\end{figure}

\begin{figure}[ht!]
	\centering
    \includegraphics[width=0.9\linewidth]{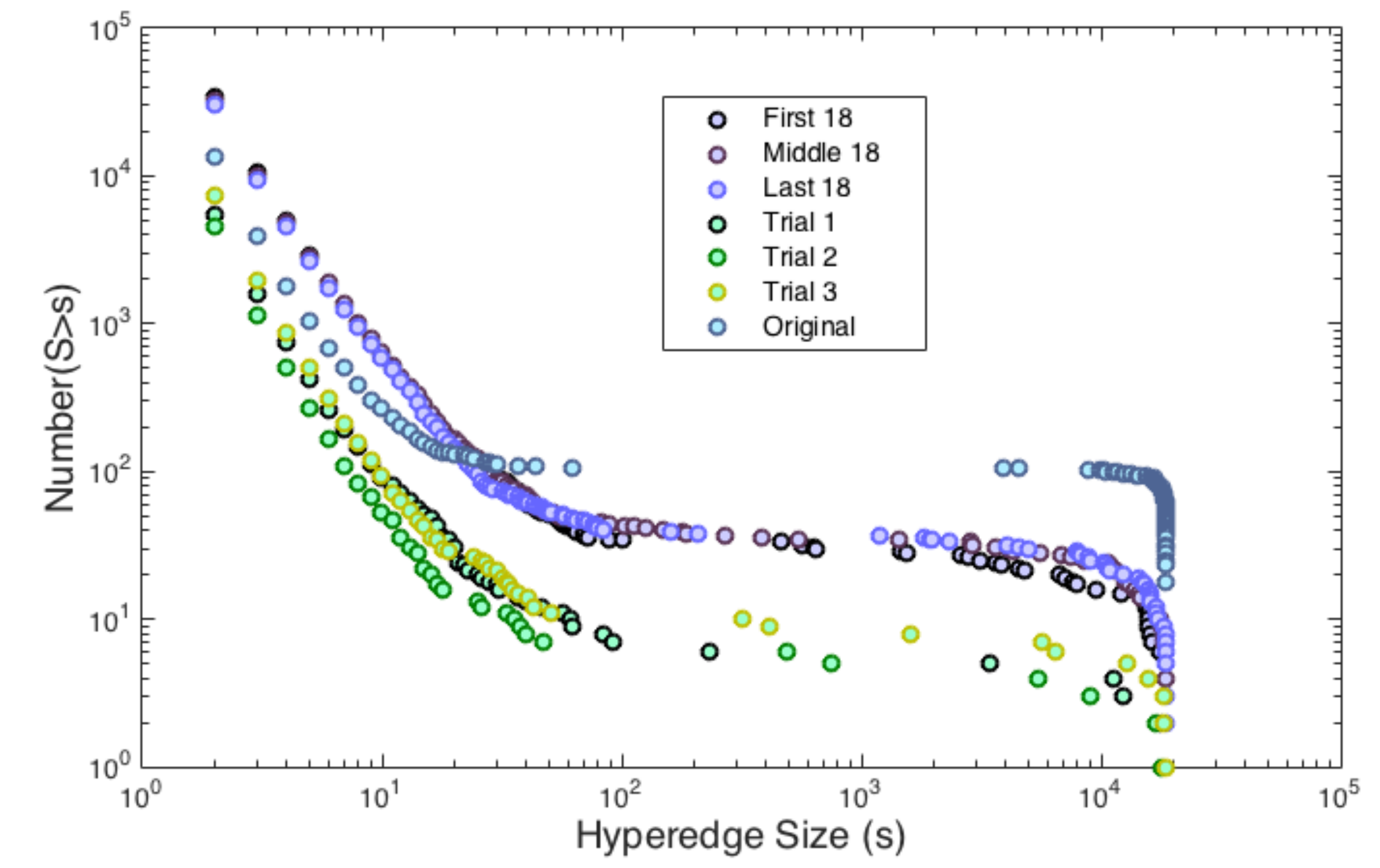}
    \caption{\textbf{Trial separation comparison:} Cumulative size distributions for two different methods for separating edge effects. In the trial-by-trial method, hypergraphs are constructed separately for each trial, while in the 18-split analysis, hypergraphs are constructed from the first, middle, or last 18 edge time series data points.}
	\label{figS2}
\end{figure}

\begin{figure}[ht!]
	\centering
    \includegraphics[width=0.9\linewidth]{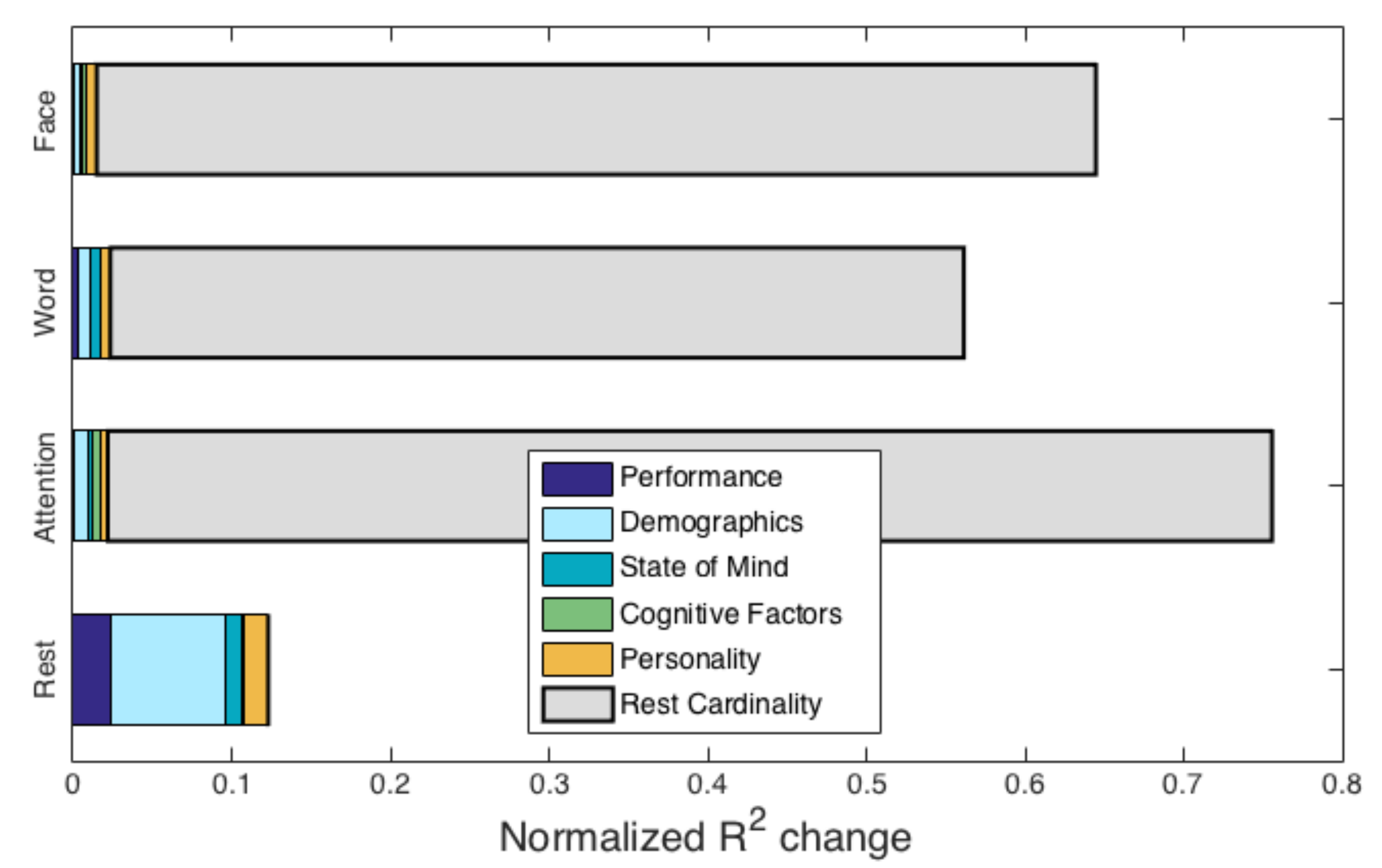}
    \caption{\textbf{Task-specific multi-task $R^2$ changes:} Normalized $R^2$ changes with respect to task-specific hypergraph cardinality for each of the four task-specific hypergraphs. Rest-specific hypergraph cardinality is included as an independent variable for the other three tasks and is the only significant predictor, which is denoted with a bold outline.}
	\label{figS3}
\end{figure}

The following information is included in this supplementary document to support the claims presented in the main work:
\begin{enumerate}
	\item A discussion of observed effects of concatenating multiple time series.
	\item Figure \ref{figS1} and Figure \ref{figS2}: Cumulative size distributions for several methods for minimizing the effect of concatenation.
	\item A discussion of the effect of time series length on hyperedge size distributions for the age-memory data set.
	\item Tables \ref{Table_All}, \ref{Table_MT}, and \ref{Table_AM}: Tables of individual difference measures grouped by category for the full analysis, multi-task data, and age-memory data.
	\item Figure \ref{figS3}: $R^2$ changes for the task-specific hypergraph cardinality regression analysis.
\end{enumerate}
\paragraph*{Methodological Considerations.}
\label{S1_Appendix}
\textbf{Edge Effects in Task Concatenation:}
In this paper, we investigate dynamic functional connectivity changes across multiple cognitive tasks and two separate imaging data sets. In order to capture changes across tasks in the multi-task data set, we concatenate the time series for all tasks, as in \cite{davison_brain_2015}. In our analysis of the age-memory data, we concatenate time series from three functional runs of the word memory task, and remove time windows from the ends of the time series of each task to reduce edge effects.  Edge effects appear to be confined to the data points adjacent to the beginning and end of each run, but we remove the full $N \times N$ adjacency matrix to ensure we are not including any edge effects in the analysis. The resulting change in the cumulative size distribution is depicted in Figure \ref{figS1}. With the edge blocks removed, there are fewer system-size hyperedges and more small hyperedges.

Figure \ref{figS1} includes a comparison with another method for treating edge effects. In this case, the time series data for each of the three tasks is filtered separately before concatenation. This approach dramatically reduces the number of hyperedges. If filtering is responsible for introducing edge effects that drive hyperedges, the number of hyperedges are likely to increase when we employ this method. Instead, only 13 subjects had non-singleton hyperedges. We choose to not analyze these results further because there are too few subjects with hyperedge data.

Two further efforts to understand the effects of concatenating across functional runs on the cumulative size distribution are depicted in Figure \ref{figS2}. In the trial-by-trial analysis, we performed the hypergraph method separately on each edge time series (10 data points each) for the three trials. Only 30 subjects have significant non-singleton hyperedges in at least one of the three trials and the number of large hyperedges is much lower than the original result. This decrease may be a result of our removal edge effects, but it is likely the shorter task length is driving the difference, as we discuss in the next \nameref{S2_Appendix} section. To explicitly investigate the effect on the size distribution caused by each transition, we also split the time series data into three sets of 18 edge time series data points. The first includes the transition between the first and second trials, the last includes the transition between the second and third trials, and the middle includes both transitions. These distributions are also plotted in Figure \ref{figS2}. We see that the overall number of hyperedges is greater than both the original age-memory hypergraph over all individuals, which is driven by a decrease in the number of system-size hypergraphs in the 18-split analysis. The distributions for all three follow similar patterns, indicating there is not a large discontinuity in the pattern of the distribution when we include both transitions.

\paragraph*{Methodological Considerations.}
\label{S2_Appendix}
\textbf{Edge Time Series Length in Hypergraph Construction:}
When we construct hypergraphs from the much shorter single task measurements within the multi-task data set, the number of large hyperedges is greatly reduced, with fewer hyperedges in the population near the system size (see Panel A of Figure \ref{fig6}). We see a similar effect when we compare the distributions seen in Figure \ref{figS2} for the split data sets. The trial-by-trial hypergraphs contain fewer hyperedges overall and far fewer system-size hyperedges than the 18-split hypergraphs. However, this increase is not driven by inclusion of the transitions alone, since the middle 18-split hypergraph contains approximately half the number of system-size hyperedges when compared to the full analysis. Since both hypergraphs are constructed across both transitions, this indicates that the edge time series length is more influential to population-level hypergraph properties than concatenation.

Further work is needed to elucidate the relationships between hyperedge size and the overall length and composition of the data set. Additionally, it remains to be determined whether there is an analogue to the scan length proposed for reliable FC estimates \cite{birn2013effect}; an edge time series length that ensures minimal fluctuations in the size distributions for longer scans. However, the very close correspondence between small-size hyperedges found during the word memory task in both data sets suggests that these hyperedges are capturing important characteristics of the dynamics within this task that are robust across imaging sessions and populations.

\begin{table}[ht!]
	\begin{tabular}{|l|l|l|l|l|}
		\hline
		Performance (Word)& Demographics & Personality & Cognitive Factors & State of Mind  \\
		\hline
		Criterion shift score & Age & PANAS (6) & OSIQ-S/O &   Arrival time \\
		Liberal Dprime & Sex & Big 5 (5) & VVQ-W/P&   Meal (hours since) \\
		Conservative Dprime & Education (years) & BIS/BAS (4) & Need for cognition &   Hours of sleep  \\
		Overall Dprime & Dominant hand  & &  SBCSQ visual  &   Physical/mental comfort   \\	
		& & & SBCSQ verbal  &  Beck Depression Inventory  \\	
		& & & Paper folding & Alcohol (Y/N)\\
		& & & Card rotation & Exercise (Y/N)\\
		 & & &  & Smoking (Y/N)\\
		 & & &  & Caffeine (Y/N) \\
		\hline
	\end{tabular}
    \caption{\textbf{Common behavioral measures in both data sets:} Categories containing measures of interest (42). For the state of mind measures, (Y/N) indicates measures where participants were asked whether they had performed the activity in the past 24 hours.}
    \label{Table_All}
\end{table}

\begin{table}[ht!]
	\begin{tabular}{|l|l|l|l|l|}
		\hline
		Performance & Demographics & Personality & Cognitive Factors & State of Mind  \\
		\hline
		Attention CS & Military rank & EPQ-R (4) & Working memory & MSW/MSF   \\
		 Face memory CS & & &  Vocabulary test & PTSD Score\\
		 Attention Dprime & &  & &  PTSD (Y/N)\\
		 Face memory Dprime & & &   & Concussion score  \\
		 & & & & Concussion 5 inventory\\
		\hline
	\end{tabular}
    \caption{\textbf{Additional behavioral measures in multi-task data:} Categories containing measures of interest. For the state of mind measures, (Y/N) indicates measures where participants were asked whether they had performed the activity in the past 24 hours.}
	\label{Table_MT}
\end{table}

\begin{table}[ht!]
	\begin{tabular}{|l|l|l|l|l|}
		\hline
		Performance & Demographics & Personality & Cognitive Factors & State of Mind \\
		\hline
		 Hit rates & Height &  Distracted &	&  Stressed (Y/N)   \\
		Failure rates & Weight & Motivated &    & Days since period  \\
		 Reaction time & Contraceptive use &  &	& Usual hours of sleep    \\	
		 &  Children (Y/N)  &  &  &  Drugs past 48h (Y/N) \\	
		 & Number of children&  &  & MMSE (dementia) \\
		\hline
	\end{tabular}
	\caption{\textbf{Additional behavioral and brain measures in age-memory data:} Categories containing measures of interest. For the state of mind activity measures, yes indicates measures where participants were asked whether they had performed the activity in the past 24 hours. Questions about daily, weekly, and monthly amounts of activity, including whether activity in the past 24 hours were more or less than usual were also recorded for all (Y/N) state of mind activities in the age-memory study.}
\label{Table_AM}
\end{table}

\section*{Acknowledgments}
We would like to thank John Bushnell for technical support. 

\nolinenumbers

\bibliographystyle{plos2015}
\bibliography{bibfile}

\end{document}